\documentclass[letterpaper,english,prb,twocolumn]{revtex4}
\usepackage[T1]{fontenc}
\usepackage[latin9]{inputenc}
\usepackage{graphicx}
\usepackage{float}

\makeatletter

\providecommand{\tabularnewline}{\\}

\@ifundefined{textcolor}{}
{%
 \definecolor{BLACK}{gray}{0}
 \definecolor{WHITE}{gray}{1}
 \definecolor{RED}{rgb}{1,0,0}
 \definecolor{GREEN}{rgb}{0,1,0}
 \definecolor{BLUE}{rgb}{0,0,1}
 \definecolor{CYAN}{cmyk}{1,0,0,0}
 \definecolor{MAGENTA}{cmyk}{0,1,0,0}
 \definecolor{YELLOW}{cmyk}{0,0,1,0}
}

\makeatother

\usepackage{babel}
\begin{document}

\title{Towards improved exact exchange functionals relying on GW quasiparticle
methods for parametrization}

\author{V. Z\'olyomi$^{1,2
}$ and J. K\"urti$^{3}$}

\address{$^{1
}$Physics Department, Lancaster University, Lancaster, United
Kingdom, LA1 4YB}

\address{$^{2
}$Wigner Research Institute, Hungarian Academy of Sciences,
P. O. B. 49, H-1525, Budapest, Hungary}

\address{$^{3
}$Department of Biological Physics, E\"otv\"os University, P\'azm\'any
P. s\'et\'any 1/A, H-1117}
\begin{abstract}
We use fully self-consistent GW calculations on diamond and silicon
carbide to reparametrize the Heyd-Scuseria-Ernzerhof exact exchange
density functional for use in band structure calculations of semiconductors
and insulators. We show that the thus modified functional is able
to calculate the band structure of bulk Si, Ge, GaAs, and CdTe with
good quantitative accuracy at a significantly reduced computational
cost as compared to GW methods. We discuss the limitations of this
functional in low-dimensions by calculating the band structures of
single-layer hexagonal BN and MoS$_{2}$, and by demonstrating that
the diameter scaling of curvature induced band gaps in single-walled
carbon nanotubes is still physically incorrect using our functional;
we consider possible remedies to this problem.
\end{abstract}
\maketitle

\section{Introduction}

Applied materials science relies heavily on the ability of theoreticians
to accurately predict the physical properties of novel materials.
Calculations are widely used as guidance in the design of new materials,
without which experiments can rapidly turn into a cumbersome
and expensive trial-and-error process. Unfortunately, the ability
of theoreticians to provide guidance to experimentalists is in practice
severely limited by the accuracy of the methods available. One particular
example for this issue is the problem of predicting the band gap of
semiconductors.

A widely used and very accurate method in the study of semiconducting
materials is density functional theory (DFT). Useful as it is, a well
known problem exists with DFT in that it severely underestimates the
band gap. This is especially true in low-dimensional materials such
as boron nitride \cite{BN_Ciraci_PhysRevB.79.115442}, polyyne \cite{Polyyne_PhysRevB.72.155420,PolyyneGW},
or even carbon nanotubes where even the diameter scaling of the curvature
induced band gap is wrong, let alone its magnitude \cite{ZolyomiKurti2004_PhysRevB.70.085403}.
This problem can be circumvented by going beyond the DFT level and
using for example the Quantum Monte Carlo (QMC) method to calculate
the band gap \cite{CASINO_0953-8984-22-2-023201}. QMC is however
extremely expensive to apply to semiconductors or insulators unless
the location of the band gap in the reciprocal space is exactly known.

\begin{figure}[b]
\includegraphics[width=0.73\columnwidth]{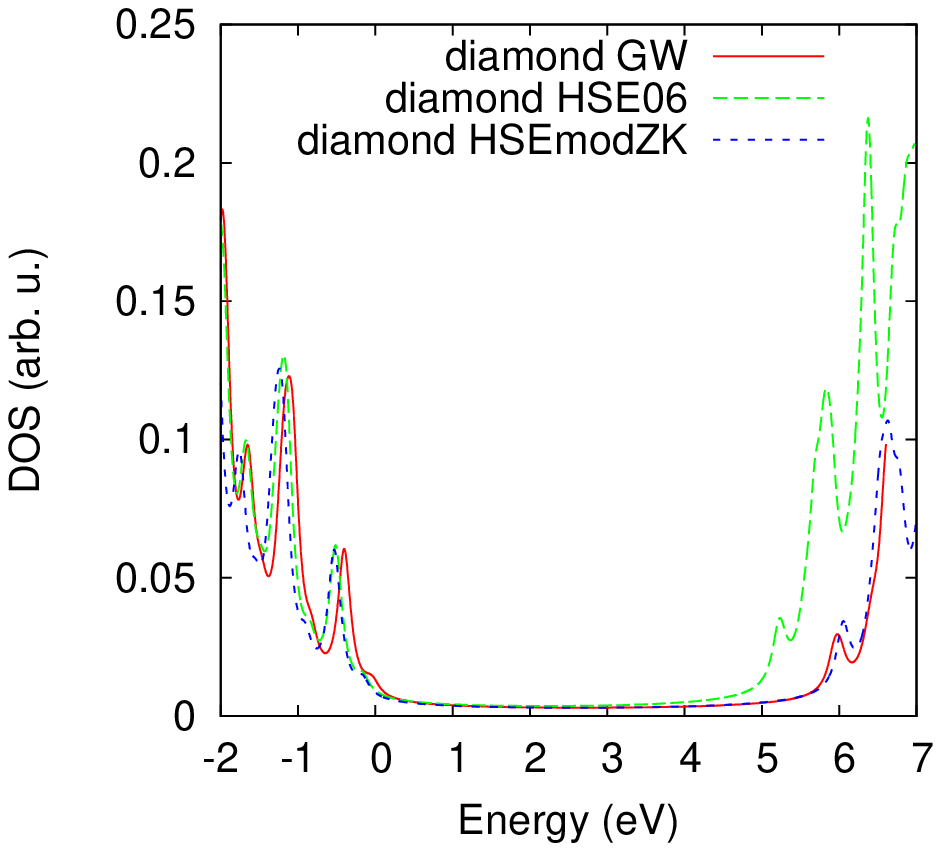}

\includegraphics[width=0.73\columnwidth]{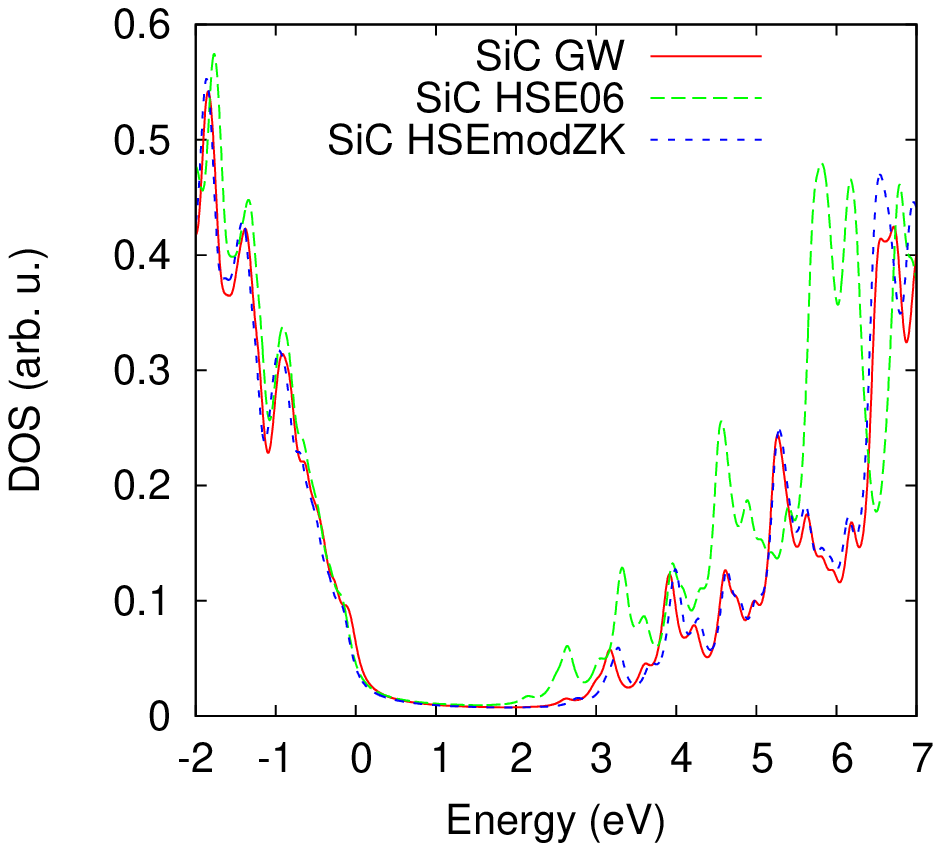}

\caption{\label{DOSfig}The electronic density of states for bulk diamond and
$\beta$ silicon carbide calculated with HSE06, the modified HSE functional,
and the GW approximation. HSE06 considerably underestimates the gap,
while our modified functional is in close agreement with the GW calculations. }
\end{figure}

Another approach for improving the band gap is the use of many-body
theory, and more specifically, the GW approximation \cite{GW_review_0953-8984-11-42-201}.
This allows one to calculate the quasiparticle band structure by starting
from the DFT wave functions and solving the Dyson equation. The resulting
band gaps are typically in very good agreement with experiments and are
a signifcant improvement over DFT results (see Figure \ref{DOSfig}). The
disadvantage of the GW approximation is that it is computationally
very expensive. While today it is possible to use the method on systems
composed of as much as a hundred atoms on massively parallel machines
\cite{BerkeleyGW_Deslippe20121269}, there are plenty of calculations
that are simply not feasible with GW to this day. Examples include
large supercell calculations such as surface intercalation, or the
study of composite materials and heterostructures, which are of particular
interest to the scientific community \cite{Britnell24022012,Britnell14062013,6384726}.
Furthermore, in low-dimensional materials the GW approach converges
extremely slowly as pointed out in a recent publication \cite{GW_MoS2_PhysRevLett.111.216805},
making it all but infeasible for the study of nanoscale materials
at this time.

Due to the computational expense of both QMC and GW calculations,
it would be advantageous to develop a computational method that is
able to approach the accuracy of QMC or the GW approximation without
the accompanying computational expenses. The trivial choice would
be the use of exact exchange density functionals, also known as hybrid
density functionals. These are known to predict a much more accurate
band gap than traditional density functionals such as the local density
approximation (LDA) or the semilocal Perdew-Burke-Ernzerhof (PBE)
generalized gradient approximation (GGA) as well as other GGA functionals.
While exact exchange functionals still underestimate the band gap,
and they in general suffer from serious problems in terms of predictive
power as pointed out recently \cite{HybridGapsUnreliable_PhysRevLett.107.216806},
in principle it is possible to modify the parameters of these functionals
in order to deliver a better accuracy when calculating the band structure.

Screened exact exchange functionals such as the Heyd-Scuseria-Ernzerhof
(HSE) functional include dielectric screening \cite{HSE06PaperJCP}
which makes them ideal for use in periodic structures including three
dimensional solids and low-dimensional materials like carbon nanotubes.
These functionals offer two easily accessible parameters that can
be adjusted without tampering with the core of the functional: these
are the amount of exact exchange, and the screening parameter. Adjusting
these parameters is possible using a variety of criteria, including
what has been dubbed the HSE12 functional \cite{HSE12PaperJCP}. For
the purposes of accurate band structure calculations it is sensible
to set the criteria to achieve the best possible match between the
band structures predicted by the modified HSE functional and GW many-body
theory.

In this work we present a modification to the HSE functional with
the help of fully self-consistent GW many-body calculations. As we
show below, it is possible to adjust the two parameters in the HSE
functional such that the band structures calculated with the modified
functional are in very close agreement with GW calculations. We compare
our modified HSE functional to the HSE12 functional. We argue that
our method makes quantitatively accurate first-principles band structure
calculations on large systems feasible. The method is of course not
without its own limitations, which we discuss at the end of the paper.

\section{The modified HSE functional}

In our calculations we rely on the VASP code \cite{VASP_PhysRevB.54.11169},
using a plane-wave basis set. Our goal is to develop an exact exchange
functional that reproduces the GW quasiparticle band gaps in semiconductors.
We use the Heyd-Scuseria-Ernzerhof 2006 (HSE06) functional as the
basis for this approach, which describes the exchange part of the
density functional as 25\% exact exchange and 75\% PBE exchange. The
modified functional differs from the original HSE06 in two parameters
only, the amount of exact exchange and the screening parameter. The
new parameters are obtained by performing a least squares fit between
the band structures obtained with our functional and GW band structures,
then changing the parameters until the root mean square (RMS) is found
to be minimal.

The fitting procedure is performed using the highest valence and lowest
conduction bands in the entire Brillouin zone on a finite, $12\times12\times12$
$k$-point grid. At each $k$-point the squared energy difference
is taken between the GW and the modified HSE valence band, then summed
up over the entire grid. The process is repeated for the conduction
band and added to the sum obtained from the valence band. The thus
obtained RMS value gauges the quality of the match between the modified
HSE functional and the GW calculation. The smaller the RMS the better
the parameters of the modified HSE functional.

The GW calculations are performed in the fully self-consistent approach
where both the Green's function and the dielectric screening are iteratively
updated until self-consistency is reached (a maximum of four iterations
were required). This allows for the best available accuracy within
GW many-body theory. Considering the difficulty of converging such
calculations in low-dimensional materials \cite{GW_MoS2_PhysRevLett.111.216805}
we restrict ourselves to three-dimensional crystals. In particular
we look at diamond and $\beta$ silicon carbide (3C-SiC) crystals.
The geometries are optimized using the HSE06 functional. The GW calculations
are then performed in fixed geometries. Both the structural relaxations
and the GW calculations are carefully tested for convergence with
respect to a variety of parameters. The convergence criteria used
here are an uncertainty of $<0.001$ \AA~ in the lattice constant
during the HSE06 optimizations and an uncertainty of $<0.05$ eV in
the energy gap during the GW calculations. The HSE06 optimizations
are found to reach convergence at an $8\times8\times8$ Monkhorst-pack
grid in $k$-space and 800 eV plane-wave cutoff energy, while the
GW calculations at a $12\times12\times12$ $k$-point grid, 400 eV
plane-wave cutoff, 16 electronic bands, and $N_{\omega}=48$ where
$N_{\omega}$ is the number of frequency points used in the contour
integrations in the complex plane.

With the fully converged GW band structures available we can make
an easy comparison between many-body theory and existing hybrid density
functionals. Figure \ref{DOSfig} shows the electronic density of
states for diamond and $\beta$ silicon carbide according to the GW
approximation and the HSE06 functional. As expected the HSE06 band
gap is much smaller. If, however, we modify the exact exchange ratio
and screening parameters in the HSE06 functional, it is possible to
rectify this problem.

We shall denote the two parameters as follows. Parameter $A_{EXX}$
is the amount of exact exchange in the density functional, i.e. the
total exchange in the functional is $A_{EXX}\cdot E_{x}^{exact}+(1-A_{EXX})\cdot E_{x}^{PBE}$
where $E_{x}^{exact}$ and $E_{x}^{PBE}$ are the exact and PBE exchange
terms, respectively. Parameter $\mu$ controls the amount of screening
in the functional; the dimension of $\mu$ is inverse length and it
determines the separation of long-- and short--range terms in the
exchange within HSE-type functionals, which is equivalent to applying
a semiempirical screening to long--range interactions. In our starting
point, the HSE06 functional, $A_{EXX}=0.25$ and $\mu=0.2$ \AA$^{-1}$.
In order to modify the HSE functional to fit the band gap to that
found in GW many-body theory, we explore this two-dimensional parameter
space within the range of $A_{EXX}\in[0.2,0.5]$ and $\mu\in[0.1,0.3]$
\AA$^{-1}$. The previously discussed RMS will be a function of these
two parameters, and in order to find the ideal settings for $A_{EXX}$
and $\mu$ this function must be minimized.

\begin{figure}
\includegraphics[width=0.8\columnwidth]{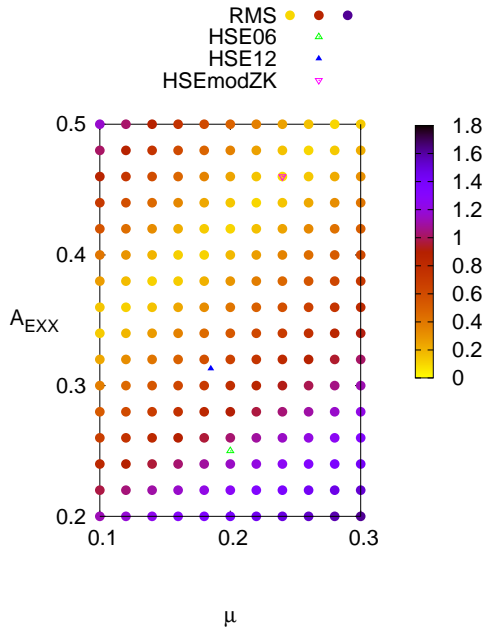}

\caption{\label{RMSfig}The root mean square of the band structure of bulk
diamond and $\beta$ silicon carbide between the modified HSE functional
and GW calculations. The minimum is found at $AEXX=0.46$ and $\mu=0.24$
\AA$^{-1}$ where the RMS is 0.10 eV; this is a substantial improvement
over HSE06 and HSE12. }
\end{figure}

Figure \ref{RMSfig} shows the RMS of the least square fit between
the GW band structures and the modified HSE functional as a function
of $A_{EXX}$ and $\mu$, averaged for diamond and $\beta$ silicon
carbide, the two benchmark materials considered. The entirety of the
topmost valence band and the bottommost conduction band is taken into
account spanning the entire Brillouin zone in a $12\times12\times12$
grid. The best fit is achieved at $A_{EXX}=0.46$ and $\mu=0.24$
\AA$^{-1}$. The figure highlights the settings of the original HSE06
functional, the HSE12 parametrization, and our modified HSE functional.
As it can be seen, the fit substantially improves between our modification
and either HSE06 or HSE12.

Now if we return to Figure \ref{DOSfig} we can see that the density
of states obtained using our modified HSE functional with $A_{EXX}=0.46$
and $\mu=0.24$ \AA$^{-1}$ agrees extremely well with the GW calculations
for both diamond and $\beta$ silicon carbide.

\section{Application of the modified HSE functional}

\begin{figure*}
\includegraphics[width=0.4\paperwidth]{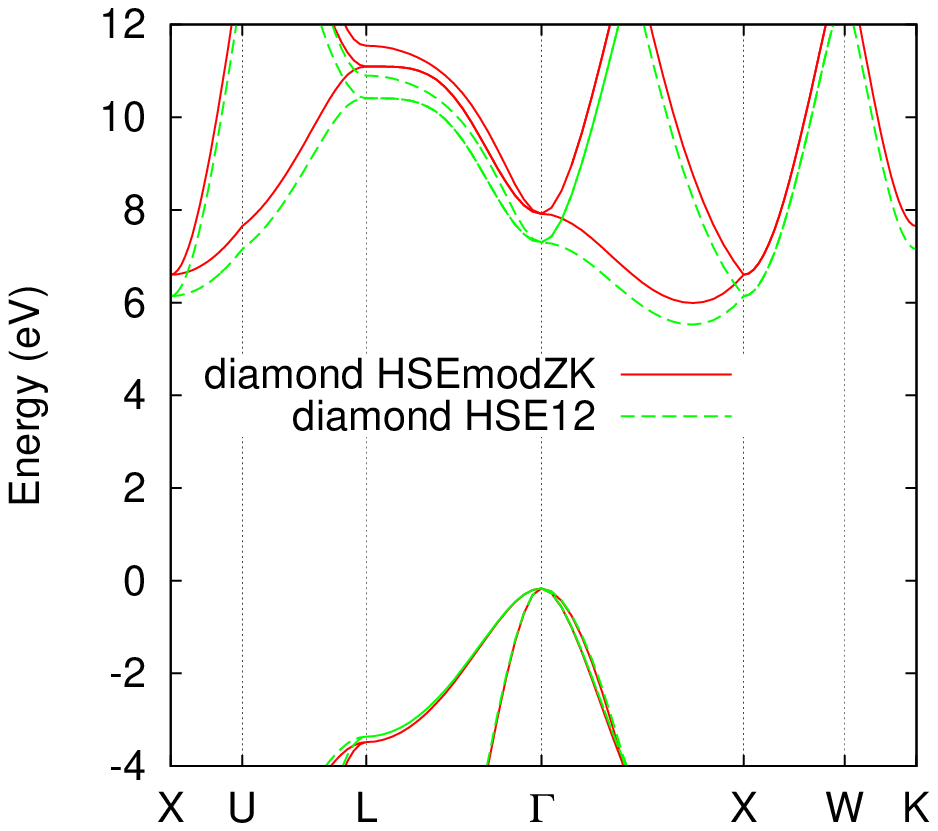}\includegraphics[width=0.4\paperwidth]{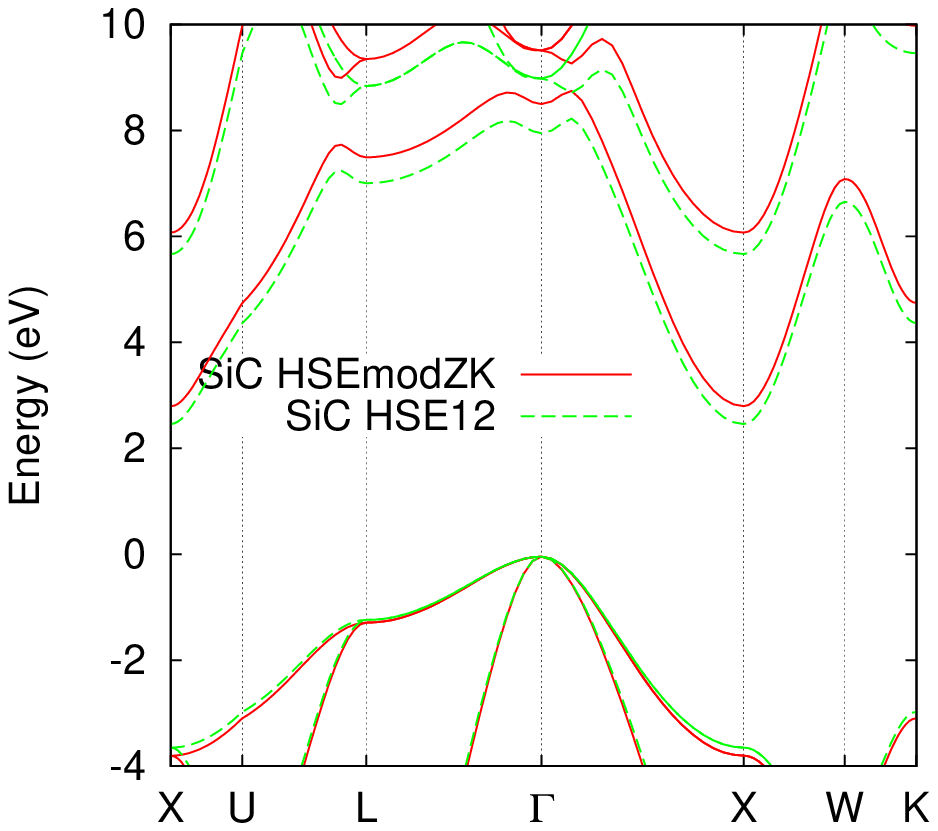}

\includegraphics[width=0.4\paperwidth]{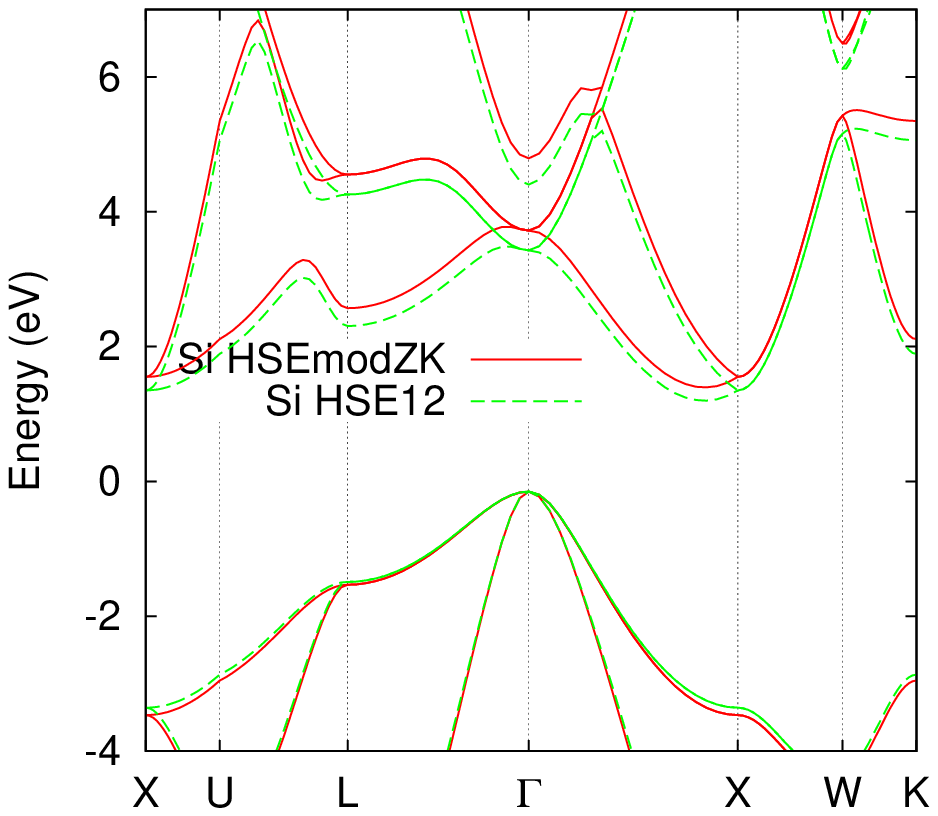}\includegraphics[width=0.4\paperwidth]{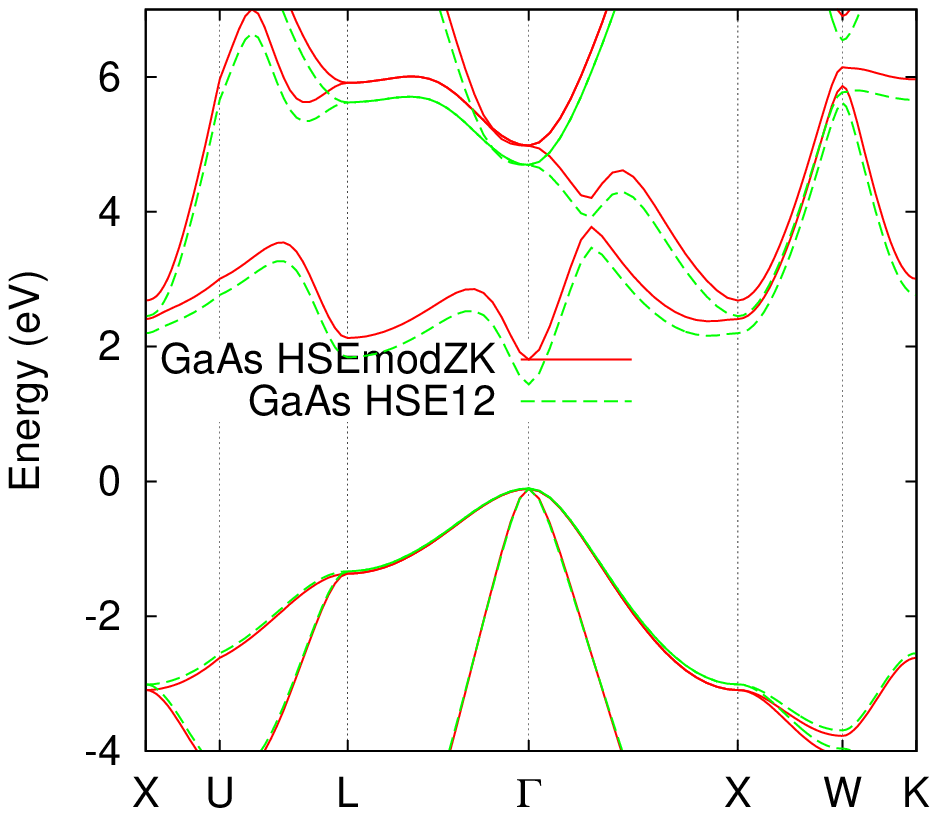}

\includegraphics[width=0.4\paperwidth]{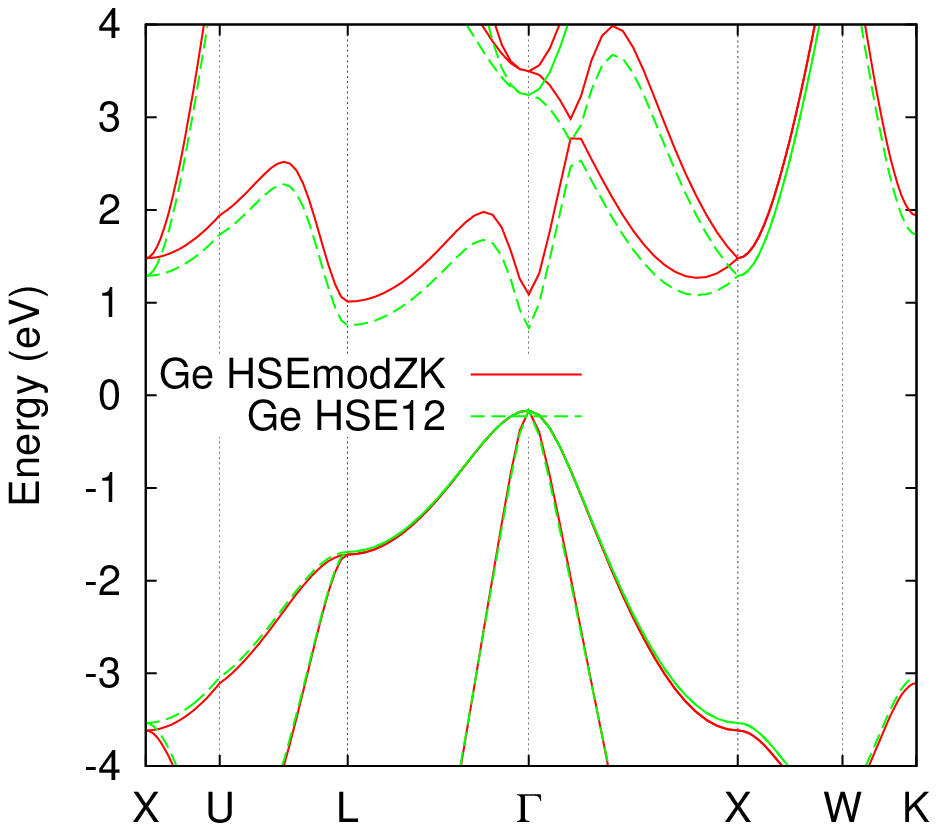}\includegraphics[width=0.4\paperwidth]{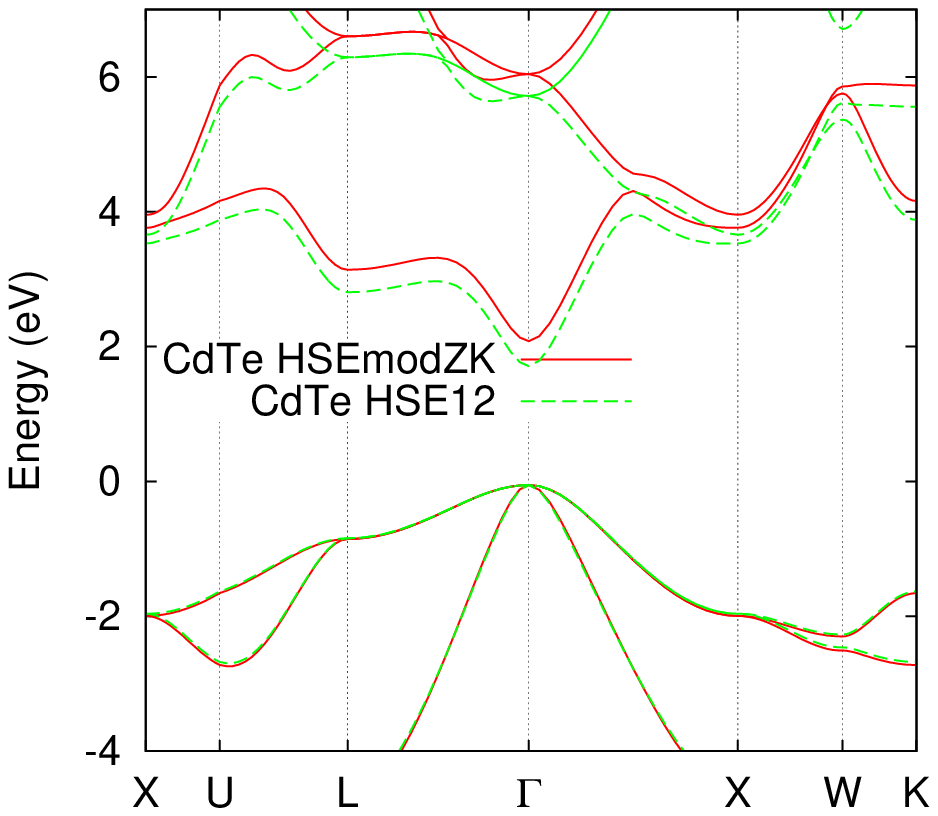}

\caption{\label{DiamondBands}The electronic band structure for bulk diamond
(top left), bulk silicon (middle left), bulk germanium (bottom left),
$\beta$ silicon carbide (top right), gallium arsenide (middle right),
and cadmium telluride (bottom right) calculated with the modified
HSE functional and compared to results obtained with the HSE12 functional.}
\end{figure*}

Now we turn our attention to the use of the modified HSE functional.
First we look at its performance on three-dimensional semiconductors,
bulk silicon (Si), germanium (Ge), gallium arsenide (GaAs), and cadmium
telluride (CdTe). Figure \ref{DiamondBands} shows the band structures
of these four materials using our modified HSE functional, along with
those of diamond and silicon carbide. Table \ref{tab:DiamondGaps}
compares their band gaps with our GW calculations. The overall agreement
is quite good, with the average absolute deviation between the band
gaps in the modified HSE functional and our GW results being a modest
0.14 eV. This is a significant improvement on the average performance
of the HSE12 functional which yields an average absolute deviation
of 0.28 eV. We must note however that in some individual cases HSE12
performs better, such as bulk silicon.

\begin{table}
\begin{tabular}{|c|c|c|c|}
\hline 
 & $\Delta{}_{HSEmodZK}$ & $\Delta_{GW}$ & $\Delta_{HSE12}$\tabularnewline
\hline 
\hline 
C & 6.16 & 6.05  & 5.69\tabularnewline
\hline 
Si & 1.55 & 1.33 & 1.35\tabularnewline
\hline 
Ge & 1.17 & 1.07 & 0.89\tabularnewline
\hline 
SiC & 2.84 & 2.70 & 2.50\tabularnewline
\hline 
GaAs & 1.92 & 2.12 & 1.54\tabularnewline
\hline 
CdTe & 2.14 & 2.08 & 1.76\tabularnewline
\hline 
\end{tabular}

\caption{\label{tab:DiamondGaps}Band gaps (in eV) for the materials in Figure
\ref{DiamondBands} using our modified HSE functional ($\Delta{}_{HSEmodZK}$)
and our GW calculations ($\Delta_{GW}$). Band gaps obtained using
the HSE12 functional ($\Delta_{HSE12}$) are provided for comparison.
Note that spin-orbit coupling is neglected here, which is strictly
speaking an important factor in CdTe.}
\end{table}

We must note here that we use geometries obtained using the HSE06
functional for all our calculations. We do this because we cannot
expect the modified functional to correctly predict the total energy
or the Hellmann-Feynman forces. Density functionals are typically
optimized to possess accurate predictive power for a broad range of
physical properties, whereas our modified HSE functional is made specifically
for the purposes of calculating accurate band structures in a fixed
geometry. This does not necessarily mean that the HSE06 geometries
will be good enough, as for example the diamond lattice constant is
notably underestimated; this is discussed below.

\begin{figure}
\includegraphics[width=0.8\columnwidth]{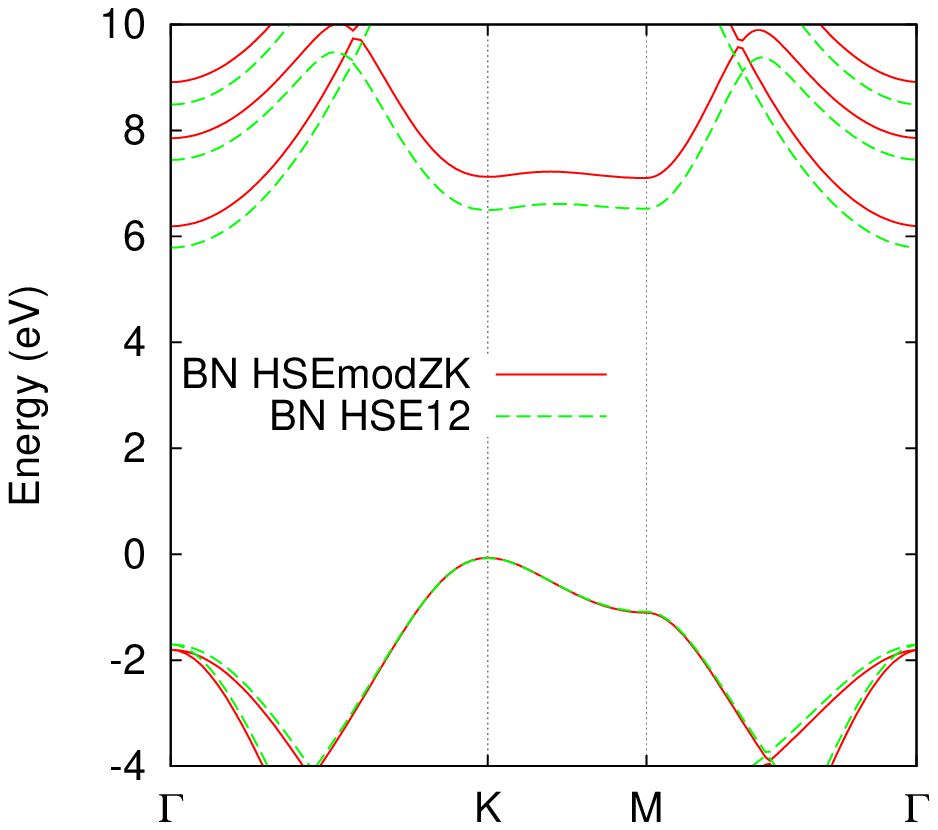}

\includegraphics[width=0.8\columnwidth]{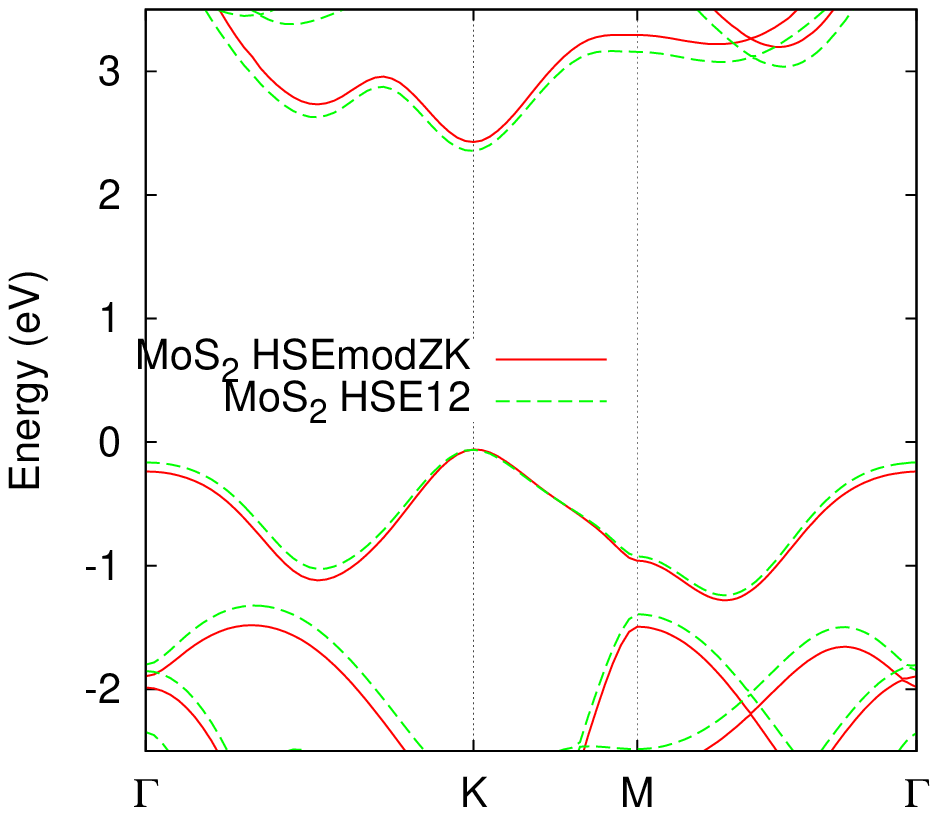}

\caption{\label{BN_MoS2_Bands}The electronic band structure for single-layer
hexagonal boron nitride (top) and molybdenum disulphide (bottom) calculated
with the modified HSE functional and compared to results obtained
with the HSE12 functional.}
\end{figure}

Now we turn to lower dimensional materials, starting with two-dimensional
nanosheets. In particular we consider single layer molybdenum disulphide
(MoS$_{2}$) and hexagonal boron nitride (h-BN). Their band structures
are shown in Figure \ref{BN_MoS2_Bands}, their band gaps are compared
to previous GW literature in Table \ref{tab:BN_MoS2_Gaps}. 

\begin{table}
\begin{tabular}{|c|c|c|c|}
\hline 
 & $\Delta{}_{HSEmodZK}$ & $\Delta_{GW}$ & $\Delta_{HSE12}$\tabularnewline
\hline 
\hline 
BN & 6.26 & 7.0 \cite{BNpaperDrummond2014} & 5.86\tabularnewline
\hline 
MoS$_{2}$ & 2.49 & 2.84 \cite{GW_MoS2_PhysRevLett.111.216805} & 2.42\tabularnewline
\hline 
\end{tabular}

\caption{\label{tab:BN_MoS2_Gaps}Band gaps (in eV) for monolayer hexagonal
boron nitride and molybdenum disulphide using our modified HSE functional
($\Delta{}_{HSEmodZK}$) and GW calculations ($\Delta_{GW}$) in the
literature. Band gaps obtained using the HSE12 functional ($\Delta_{HSE12}$)
are provided for comparison.}
\end{table}

Our band gap for MoS$_{2}$ is 2.49 eV which compares reasonably to
literature estimates of 2.50 eV and 2.66 eV quasiparticle gaps \cite{MoS2_GW_PhysRevB.87.155304}
obtained using the single shot G$_{0}$W$_{0}$ and the partially
self-consistent GW$_{0}$ method, respectively, but is notably below
the fully self-consistent GW band gap of 2.84 eV \cite{GW_MoS2_PhysRevLett.111.216805}
(albeit the experimental lower bound for the quasiparticle band gap
is 2.5 eV \cite{MoS2_exp_gap_2014}). The band gap of h-BN is found
to be 6.26 eV here, which falls between recent predictions for the
band gap \cite{BNpaperDrummond2014} in HSE06 (5.65 eV) and single-shot
G$_{0}$W$_{0}$ calculations (7.0 eV). While our functional significantly
improves the band gap for both MoS$_{2}$ and h-BN, it still underestimates
the GW results. This suggests that care must be taken when applying
our functional for the study of two-dimensional materials.

Finally we consider a more difficult problem, that of the curvature
induced band gaps in carbon nanotubes. In single-walled carbon nanotubes
(SWCNTs) the simple zone-folding picture predicts that whenever the
difference between the chiral indices $(n,m)$ is a multiple of 3,
the nanotube is metallic. In reality, the finite curvature of the
SWCNT surface induces a band gap in all cases except for armchair
$(n,n)$ SWCNTs. This means that $(n,0)$ zigzag nanotubes exhibit
a small, curvature induced band gap when $n=3k$. This has been known
for many years, and it has been measured in STM \cite{STM_gaps_SWCNT_Ouyang27042001}
which has shown that the curvature induced band gap approximately
scales as $1/d^{2}$ where $d$ is the diameter of the SWCNT. Indeed,
being a curvature induced effect, this gap must scale with an even
power of $1/d$. DFT calculations however are unable to correctly
reproduce this behavior. The LDA in fact predicts a $1/d^{3}$ scaling
\cite{ZolyomiKurti2004_PhysRevB.70.085403}, which is a clear indication
of the failure of semilocal density functional theory to accurately
describe electronic excitations. Hence we use our modified HSE functional
to recalculate the band structure of zigzag nanotubes in an effort
to rectify this failing of semilocal DFT functionals.

\begin{figure}
\includegraphics[width=0.8\columnwidth]{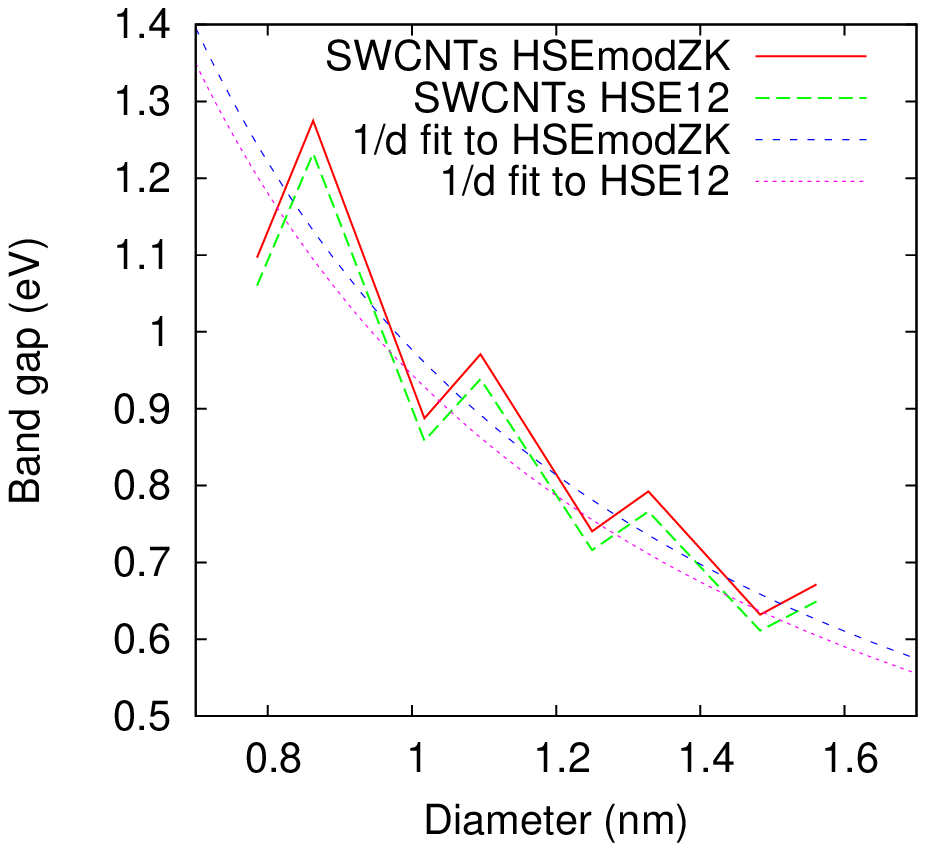}

\includegraphics[width=0.8\columnwidth]{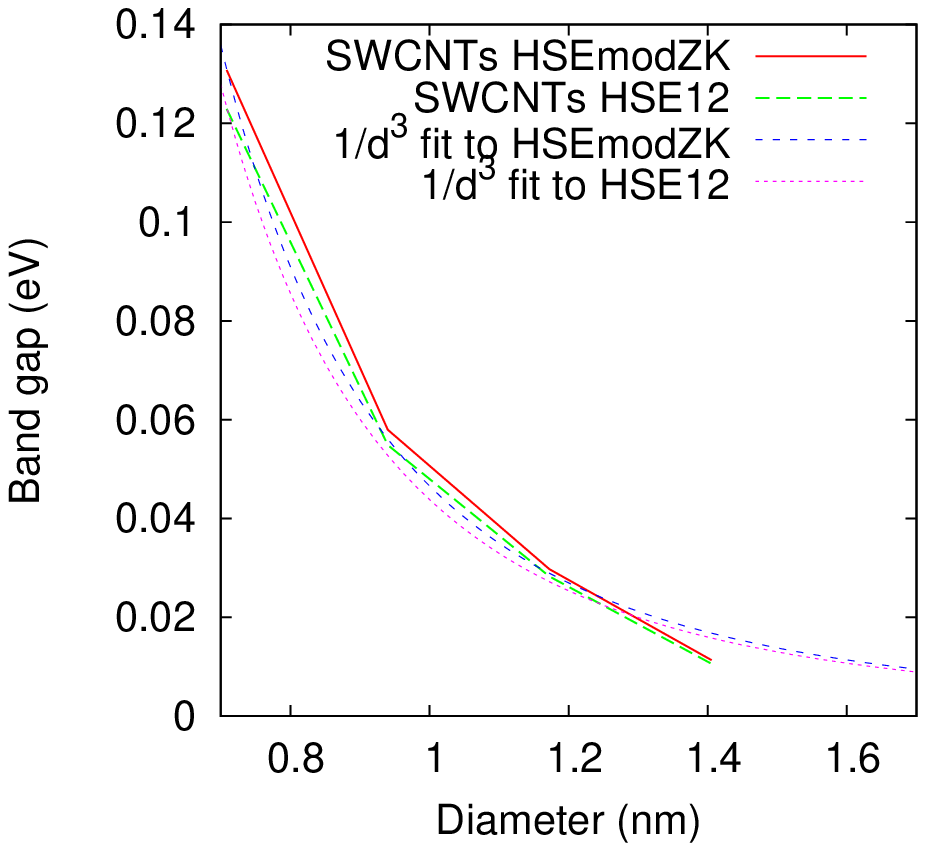}

\caption{\label{CNT_gaps}The primary (top) and secondary (bottom) band gaps
of SWCNTs according to our modified HSE functional and the HSE12 functional.
The primary gaps correctly scale with the inverse diameter ($E_{gap}=a/d$
where $a_{HSE12}=9.44$ eV$\cdot$\AA~ and $a_{HSEmodZK}=9.76$ eV$\cdot$\AA;
compare with the LDA result of $a_{LDA}=7.1$ eV$\cdot$\AA) and
the previously reported buckling caused by the trigonal warping effect
is also observable. The curvature induced secondary gaps follow a
$1/d^{3}$ scaling ($E_{gap}=b/d^{3}$ where $b_{HSE12}=43.8$ eV$\cdot$\AA$^{3}$
and $b{}_{HSEmodZK}=46.6$ eV$\cdot$\AA$^{3}$; compare with the
LDA result of $b{}_{LDA}=34.1$ eV$\cdot$\AA$^{3}$), which is unphysical
considering that curvature effects should scale with an even power
of $1/d$.}
\end{figure}

Figure \ref{CNT_gaps} shows the band gap of zigzag nanotubes from
$(9,0)$ to $(20,0)$, with the curvature induced gaps plotted separately
in the lower panel, using both our modified HSE functional and the
HSE12 functional. As it can be seen the diameter scaling of the curvature
induced band gap still follows the $1/d^{3}$ scaling found in LDA
calculations. This illustrates that the modified HSE functional is
unfortunately not entirely suited for one-dimensional materials, and
neither is HSE12. On the other hand, both functionals correctly capture
the physics in the primary gaps of semiconducting zigzag nanotubes,
as both the $1/d$ scaling of the gap and the trigonal warping related
buckling effect are correctly described. The band gaps are larger
in our modified HSE functional, as expected.

\section{Discussion}

We have demonstrated above that empirically fitting the parameters
of HSE-type exact exchange functionals to GW calculations is a feasible
and sound approach to improve the performance of such functionals
in band structure calculations. We have in particular demonstrated
that by changing the parameters of the HSE06 functional it is possible
to calculate band structures that agree with GW calculations for the
valence and conduction band with an average RMS error of 0.1 eV, which
is very impressive for a density functional.

The importance of this finding cannot be stressed enough. A very simple
modification enables one to use an HSE-type density functional to
closely approach the accuracy of GW. Since HSE-type calculations are
not only significantly cheaper than GW but are also much easier to
converge, the modified HSE functional can be used to give accurate
theoretical predictions for the band structure of the kind of complex
materials where GW is simply not feasible. Therefore, a correctly
parametrized modification of the HSE functional is invaluable for
condensed matter research, and we have demonstrated here that such
a parametrization is possible to find.

However, it is important to be aware of two shortcomings of this study.
First of all, the band gaps obtained in our GW calculations are consistently
higher than expected; this is especially notable in the case of diamond
where our GW band gap is 6.05 eV while experimentally it is 5.47 eV.
Previous G$_{0}$W$_{0}$ calculations performed on the experimental
geometry yielded a much better match with a band gap of 5.68 eV \cite{DiamondGW_PhysRevB.45.8239}.
Similarly, our GW band gaps for Si, Ge, and GaAs are notably larger
than previous GW calculations \cite{SiGeGaAsGW_PhysRevB.69.125212}.
Considering that we rely on fully self-consistent GW rather G$_{0}$W$_{0}$
or GW$_{0}$, this is a surprising result. One must bear in mind,
however, that electron-phonon coupling introduces a notable band gap
renormalization in diamond lattices. In particular, in diamond itself
the gap is reduced by 0.4 eV \cite{DiamondGW_PhysRevB.45.8239,EPC_gaprenorm_PhysRevB.89.214304,EPC_T4_2014},
which accounts for most of our overestimation and leaves only an error
of 0.18 eV unexplained.

This remaining difference may be due to a number of reasons. One may
be the use of HSE06 geometries instead of experimental lattice constants.
The lattice constant of diamond is slightly underestimated at a value
of 3.54 ~\AA~ in our calculations as opposed to the experimental
value of 3.57 ~\AA. However, the calculated lattice constant of silicon
is 5.44 ~\AA~ which matches the experimental 5.43 ~\AA~ rather well,
so the lattice constant may not be the biggest factor. Much more likely
is that our calculations face convergence issues. This is a surprising
finding since, as mentioned earlier in the paper, we have performed
a full convergence test with respect to all relevant parameters in
the calculation, and we have done so without separating them (that
is, they were not treated independently). We have used direct energy
gaps at high symmetry points in the Brillouin zone as the convergence
criteria. Our finding was that these energies change less than $0.05$
eV when we try to exceed the converged parameter set.

One possibility is that the calculations have reached a local minimum
in the parameter space of the GW parameters, and that we need to go
much further for true convergence. In particular, it has been suggested
that the number of empty bands should be on the order of a few hundred
in diamond lattice materials such as silicon \cite{SiGeGaAsGW_PhysRevB.69.125212}.
On the other hand, the expectation is that the band gap will approach
the converged value from below, while in our case the gap is overestimated.
The reason for this discrepancy is currently unclear, although the
next step is obviously to extend the convergence tests in the GW calculation
to greater ranges in the parameter space. This, however, goes well
beyond the scope of this work, since the calculations are significantly
more expensive to perform if for example the number of bands taken
into account is increased to a few hundred.

We emphasize that these potential convergence problems in our GW calculation
do not impact the main message of our paper, namely that it is perfectly
possible to fit an HSE-type exact exchange functional to GW calculations.
The parameters of $A_{EXX}=0.46$ and $\mu=0.24$ \AA$^{-1}$ will
no doubt need to be further adjusted before the modified HSE functional
suggested here can be used for quantitative predictions of the band
gap in semiconductors. In addition, in order to determine a final
set of parameters to use in the modified HSE functional it would be
advisable to perform the calculations over a large set of materials
comprising lattices of different symmetries and atomic composition.
The parameters in this paper should be used only as guidelines for
future calculations.

There is a second issue with the modified HSE functional, that of
its performance in low-dimensional materials. As our GW calculations
seem to overestimate the gap in three-dimensional semiconductors,
it is somewhat surprising to find that the modified HSE functional
underestimates the band gap in the case of low-dimensional structures.
The band gaps of MoS$_{2}$ and monolayer boron nitride are both too
small and the diameter scaling of curvature induced secondary gaps
in SWCNTs is the same unphysical $1/d^{3}$ dependence as was found
in LDA. One possible reason for this is that our functional is parametrized
exclusively on three-dimensional crystals due to the difficulty in
converging GW calculations in low dimensions. Strictly speaking the
dielectric screening in a low-dimensional material is expected to
be substantially different from that observed in three dimensions,
hence in low-dimensional materials the screening parameter $\mu$
is expected to be different. As such, the finding that our modified
HSE functional faces difficulties in low dimensions is not surprising.

The solution to the above problem is of course a reparametrization
of the modified HSE functional for use in two-- and one--dimensional
materials. This requires a number of fully converged and fully self-consistent
GW calculations to be performed on low-dimensional structures to act
as benchmarks. These are very difficult to obtain as we have seen
in the case of polyyne \cite{PolyyneGW} and boron nitride \cite{BNpaperDrummond2014},
and our current computational resources prohibit us from performing
them. Nevertheless the relative success of our modified HSE functional
in three dimensions indicates that this reparametrization is the next
logical step towards improved exact exchange density functionals for
use in electronic band structure calculations.

Finally we must note that, considering recently reported performance
issues in exact exchange functionals \cite{HybridGapsUnreliable_PhysRevLett.107.216806},
it is quite possible that such density functionals cannot accurately
predict the curvature induced gaps of SWCNTs at all. The qualitative
agreement between the HSE12 functional and our modified HSE functional
in the case of the curvature induced gaps of SWCNTs further suggests
that this may be the case. In order to convincingly answer this question
it would be necessary to not only reparametrize the modified HSE functional
for one-dimensional materials, but also to perform fully self-consistent
GW calculations on the relevant SWCNTs, which are computationally
highly demanding due to known convergence issues in low dimensions.

\section{Conclusions}

We have presented an empirical reparametrization of the HSE06 exact
exchange density functional by fitting the amount of exact exchange
and screening in the functional in order to achieve the closest match
possible to GW calculations. We have demonstrated that this method
is sound and accurate, as our modified HSE functional performs very
well in three-dimensional materials. Since the modified HSE functional
is a significantly cheaper method than the GW approximation, being
both a better scaling and -- more importantly -- a faster converging
method, our functional can be used to perform calculations on the
kind of large systems that cannot be feasibly treated in GW while
approaching the accuracy of GW at the same time. We have discussed
the limitations of the modified HSE functional in low-dimensional
materials and outlined the next steps towards the development of accurate
exact exchange density functionals for the study of the electronic
structure of nanoscale materials.

\section{Acknowledgments}

Work supported by OTKA grant K81492. We thank Dr. Andor Korm\'anyos
for valuable discussions.


\end{document}